\documentclass[sigconf]{acmart}

\usepackage{hyperref}
\usepackage{microtype}
\usepackage{parskip}
\usepackage{xcolor}

\title[Procedural Collapse: A Structural Account of Disengagement in LLM-Assisted Writing]{\textit{Procedural Collapse}: A Structural Account of Disengagement\\in LLM-Assisted Writing}

\author{JaeWon Kim}
\affiliation{%
  \institution{University of Washington}
  \city{Seattle}
  \country{USA}}
\email{jaewonk@uw.edu}

\author{Katelyn X. Mei}
\affiliation{%
  \institution{University of Washington}
  \city{Seattle}
  \country{USA}}
\email{kmei@uw.edu}

\copyrightyear{2026}
\acmYear{2026}
\setcopyright{cc}
\setcctype{by}
\acmConference[UIST Adjunct '26]{The 39th Annual ACM Symposium on User Interface Software and Technology}{November 02--05, 2026}{Detroit, MI, USA}
\acmBooktitle{The 39th Annual ACM Symposium on User Interface Software and Technology (UIST Adjunct '26), November 02--05, 2026, Detroit, MI, USA}
\acmDOI{10.1145/3830397.3841789}
\acmISBN{979-8-4007-2855-6/2026/11}

\begin{document}

\begin{abstract}
When students use large language models for writing, the dominant explanation for disengagement is dispositional: they are over-reliant, and the remedy is to scaffold self-regulation. We argue that a structural explanation is needed, offering an alternative basis for design interventions to support appropriate AI-assisted writing. Current LLM writing interfaces induce \textit{procedural collapse}: the replacement of an iterative, self-paced writing process with a single output that shifts the writer's task from generation to comprehensive evaluation. Because that evaluation is costly, shallow engagement becomes the default, and the cognitive work writing was supposed to produce goes unperformed. The framework points toward design directions that reduce the burden on writers to self-regulate, including decomposed interaction, goal elicitation as a default first step, and single-level output. They complement metacognitive scaffolding by restructuring the interaction itself.
\end{abstract}

\maketitle

\section{Introduction}
Research on LLM-assisted writing tends to center the problem around users' behaviors. Students over-rely on AI output~\cite{parasuraman2010complacency}, offload thinking they should do themselves~\cite{risko2016cognitive}, or lack the skill to use LLMs well~\cite{fan2024metacognitive}. Users often accept AI-generated output with little scrutiny. But the fixes these accounts suggest, such as reflection prompts, review nudges, and added friction, all lean on the same self-regulation that is weakest in the people most at risk, including adolescents and writers with limited executive function~\cite{diamond2013executive}.

We argue that cognitive disengagement with AI-generated outputs is structural, not dispositional. LLM writing tools induce \emph{procedural collapse}: they compress a sequential writing process into a single finished draft, turning the writer's task from producing text into evaluating all of it at once, under conditions that make skimming the default. Without an LLM, a writer handles a text in stages: sketching a structure, drafting a paragraph, catching a mismatch, revising, then stepping back and iterating. An LLM folds that sequence into one output, which still needs attention at every level, all at once. It may need more scrutiny than a writer's own draft: the text can make stylistic choices the writer would not have made, misread the writer's intent, or introduce ideas the writer never considered, all of which must be caught after the fact.

This process is not the same as cognitive offloading. Offloading treats disengagement as a choice: the writer judges that thinking costs more than delegating. That story fits users who are lazy, rushed, or miscalibrated. Our claim concerns execution, not the decision. A writer can mean to check the output carefully and still lose the thread partway through because the levels arrive at once rather than one at a time. When the evaluation load exceeds working-memory capacity, scrutiny degrades rather than stops: attention narrows to the cheapest signal, surface completeness, while the levels where the actual thinking happens go unexamined. This happens even to users with motivations to critically engage, yielding a testable prediction: engagement falls as working-memory demands rises, regardless of motivation or intention. As such, an offloading account that predicts that motivation drives engagement fails to capture theses. 

\section{Background: Writing as a Multi-Level Constitutive Process}
\label{sec:background}
Writing builds understanding, and it does so through the act of composition itself. Putting ideas into prose forces their structure into the open and produces conceptual connections that were not there before the writer started~\cite{emig1977writing,galbraith1999writing,galbraith2009writing}. The effect is strongest on analytically demanding tasks: the more intellectual challenges the writing imposes on the writer, the deeper their engagement with the material~\cite{langer1987writing}. This matters for tool design because it suggests the value of writing is not in the finished output but in the process that produces it, and a tool that shortcuts the process can deliver the output while removing the value.

That value is unevenly distributed across the work. Some sub-processes carry little cognitive benefit for the writer---fixing typos, formatting citations, smoothing a transition whose meaning is already settled---while others produce the understanding writing is prized for: working out an argument by trying to state it, catching two claims in tension, discovering mid-draft that you do not yet know what you think. The gains from writing show up only when the task demands this second kind of work, elaboration and confrontation with one's own knowledge gaps, and not when it is routine transcription~\cite{klein1999reopening,bangert2004effects}. We call the first kind of work \emph{instrumental} and the second \emph{constitutive}. The distinction is what lets us say precisely what a writing tool can safely take over and what it cannot: constitutive work is epistemic, reshaping the writer's own understanding, so delegating it displaces the point of writing, while delegating instrumental work leaves that understanding intact~\cite{Kirsh1994DistinguishingActiona}.

The process unfolds in steps because attending to every level of a text at once exceeds working memory's capacity. A writer moves among planning, drafting, and reviewing, managing complexity by attending to levels in sequence rather than in parallel~\cite{Flower1981CognitiveWritingf,Kellogg2013ModelWritingo,McCutchen1996CapacityCompositiona}. This sequencing gives each level the attention needed for writing to build understanding. A process that optimizes for producing finished draft removes this process. When an LLM returns the entire text at once, the levels the writer would have addressed in sequence demand attention simultaneously, leaving no place for the cognitive work that the step-by-step process made possible.

A delegation is \emph{instrumental} when the writer retains the work that builds understanding and delegates the rest: grammar on a paragraph the writer composed, citation formatting, a settled transition. Understanding is preserved because the constitutive work is already complete. A delegation is \emph{indiscriminate} when a single prompt bundles both kinds together. ``Write an essay arguing that\ldots'' delegates the argument, the evidence, the organization, and the sentences at once, and much of that work would have built understanding had the writer performed it~\cite{klein1999reopening,bangert2004effects}. Procedural collapse occurs here: while the writer would have addressed the draft at every level, this process fails to take place because the cost of examining all of it at once, and inevitably shallow engagement becomes the default.

A third pattern preserves the writer's role in the thinking while using the LLM to sharpen it: requesting weak points in a drafted argument, counterarguments to address, or challenges to their reasoning. The LLM becomes an interlocutor, producing engagement the writer would not have reached alone. Where a single-delivery interface invites indiscriminate delegation, an interaction can instead be designed to support this kind of collaboration. These patterns are not fixed traits of a writer. The same person delegates instrumentally on one task and indiscriminately on the next, depending on what the interaction makes easy, which is why interaction structure, rather than the writer's willingness to exert effort, is the right target for design.

\section{Design Implications}
\label{sec:design}
If cognitive disengagement follows from the structure of the interaction, the response is to restructure the delivery, moving interactions away from the indiscriminate delegation induced by a single-delivery interface and toward collaboration that cognitively engages the writer in the process. Three directions follow.

\begin{enumerate}
    \item \emph{Decomposed interaction.} If all levels arrive at once, the interaction can present them in sequence rather than simultaneously: an outline the writer evaluates and revises before any prose is generated, then drafting, then revision. The writer passes through the levels one at a time, as in unassisted writing, at a scale working memory can accommodate.
    \item \emph{Goal elicitation as a first step.} Much of the thinking in writing is clarifying what one intends to argue. A writer who prompts without having done this leaves the model to supply goals implicitly, and the work of clarification goes unperformed. If the tool elicits the writer's goals and current understanding before generating text, that work becomes the natural start of the session rather than something the writer must remember.
    \item \emph{Single-level output.} Independent of any sequence, each response can stay at the level the prompt concerns rather than silently resolving the others: a prompt about structure receives structural feedback, not finished prose. The writer retains agency over the levels the LLM does not address, and the cost of evaluation stays bounded.
\end{enumerate}

The design directions above are not unprecedented. A growing body of work reconceives writing assistance beyond one-shot generation~\cite{lee2024designspace}: interfaces that elicit and represent a writer's intent before drafting~\cite{kim2025intentflow,ramu2024zooming}, outline-guided and staged-drafting tools that separate planning from prose~\cite{lee2025navigating,xiong2025beyond}, and argumentative-writing assistants that support hierarchical planning and iterative revision under author control~\cite{zhang2023visar}. What we add is not another such system but a generative framework that explains why they can work: they restructure the task so engagement happens at a scale working memory can accommodate.

\section{Conclusion: From User Deficit to Interaction Design}
\label{sec:conclusion}
The dispositional explanation identifies the behavior correctly but oversimplifies its cause. Procedural collapse names the condition behind it: a step-by-step writing process replaced by a single draft whose comprehensive evaluation is costly enough that shallow engagement prevails. Sorting delegations by what they displace, whether instrumental, indiscriminate, or collaborative, shows a designer which handoffs cost the writer their understanding, which are neutral, and which deepen it.

Under cognitive load, the mind defaults to shallow, low-effort processing; sustained scrutiny must be actively recruited~\cite{Kahneman2012Thinking}. An account that attributes disengagement to the user treats this default as a personal shortfall and places the burden on writers to override it through self-regulation. That burden is heaviest for those least able to meet it, including children and writers with ADHD whose executive function is still developing or already strained. Locating the failure in the interaction changes the available response: a writer cannot readily override a cognitive default, but the interaction that triggers it can be redesigned and tested. Treating cognitive disengagement as an interaction problem is therefore more generative than treating it as a user deficit: the latter yields little beyond prompts to self-regulate, while the former opens a design space and a set of testable predictions.

\begin{acks}
JaeWon Kim would like to acknowledge the CERES Network, University of Washington Global Innovation Funds (GIF), and Student Technology Funds (STF), which provided support for this work.
\end{acks}

\bibliographystyle{ACM-Reference-Format}
\bibliography{references}

\end{document}